\documentclass[10pt,twocolumn,twoside]{IEEEtran}
\usepackage{graphicx,amsmath,amsfonts}

\def\QED{\quad\vbox{\hrule \hbox{\vrule\kern3pt
      \vbox{\kern3pt{}\kern3pt}\kern3pt\vrule}\hrule}}

\newcommand{\RR}{{\mathbb{R}}}

\newtheorem{Tm}{Theorem}[section]

\newtheorem{Cr}[Tm]{Corollary}

\newtheorem{re}[Tm]{Remark}

\numberwithin{equation}{section}

\begin{document}

\bibliographystyle{plain}

\title 
{Sparse approximation property and  stable recovery of sparse signals from noisy measurements}


 \author{Qiyu Sun 
\thanks{Copyright (c) 2011 IEEE. Personal use of this material is permitted. However, permission to use this material for any other purposes must be obtained from the IEEE by sending a request to pubs-permissions@ieee.org}
\thanks{Q. Sun is with the Department of
 Mathematics,  University of Central Florida,
Orlando, FL 32816, USA. His telephone number is 407-823-4839, fax number is 407-823-6253, and email address is qiyu.sun@ucf.edu}
}


\maketitle


\date{\today}




\begin{abstract}
 In this paper,  we introduce  a
 sparse approximation property  of order $s$ for a measurement matrix ${\bf A}$:
$$\|{\bf x}_s\|_2\le D \|{\bf A}{\bf x}\|_2+ \beta  \frac{ \sigma_s({\bf x}) }{\sqrt{s}}
\quad {\rm for\ all} \ {\bf x},$$
where ${\bf x}_s$ is the best $s$-sparse approximation of the vector ${\bf x}$ in $\ell^2$,
 $\sigma_s({\bf x})$ is the $s$-sparse approximation
error of the vector ${\bf x}$ in $\ell^1$, and $D$ and $\beta$ are positive constants.
 The sparse approximation property for a measurement matrix can be thought of as a weaker version
of its restricted isometry property and a stronger version of its null space property.
In this paper, we show that the sparse approximation property  is an appropriate condition on a measurement matrix
to consider  stable recovery of any compressible signal from its noisy measurements. In particular, we show that
  any compressible signal
can be  stably  recovered  from its noisy measurements via solving an $\ell^1$-minimization problem
if the measurement matrix  has the sparse approximation  property with $\beta\in (0,1)$, and conversely
the measurement matrix  has the sparse approximation property with $\beta\in (0,\infty)$ if  any compressible signal
can be  stably  recovered  from its noisy measurements via solving an $\ell^1$-minimization problem.

\end{abstract}
\maketitle


\section{Introduction}
\setcounter{equation}{0}

Given  positive integers $m$ and $n$ with $m\le n$ and a measurement matrix ${\bf A}$ of size $m\times n$, we
consider the  problem of compressive sampling in recovering a compressible signal
${\bf x} 
\in {\mathbb R}^n$ from its noisy measurements
${\bf z}={\bf A} {\bf x}+{\bf n}
$
via solving the  following
$\ell^q$-minimization problem:
\begin{equation}\label{minimizationproblem}
\min \|{\bf y}\|_q^q \quad {\rm subject\ to} \ \|{\bf A}{\bf y}-{\bf z} \|_p\le \epsilon,
\end{equation}
where $0<q\le 1, q\le  p\le \infty, \epsilon\ge 0$, and
 the measurement noise ${\bf n}$ satisfies
$\|{\bf n}\|_p\le \epsilon$
(\cite{candestao05} -- \cite{Daubechies10}).
Here $\|\cdot\|_q, 0<q\le \infty$, stand for the  ``$\ell^q$-norm" on the Euclidean space.

%

Given a subset $S\subset\{1, \ldots, n\}$  and a vector ${\bf x}\in {\mathbb R}^n$, denoted
 by ${\bf x}_S$ the vector whose components on $S$ are the same as those of the vector ${\bf x}$ and
vanish on the complement $S^c$.
A vector ${\bf x}\in {\mathbb R}^n$
is said to be {\em $s$-sparse} if ${\bf x}={\bf x}_{S}$ for some subset
$S\subset\{1, \ldots, n\}$ with its cardinality $\# S$  less than or equal to $s$, where $s\ge 1$.
Denote by $\Sigma_s$ the set of all $s$-sparse vectors.
Given a vector ${\bf x}$, its best $s$-sparse approximation vector ${\bf x}_s$ in $\ell^q$
 is an $s$-sparse vector which has minimal distance to ${\bf x}$ in $\ell^q$; i.e.,
$ \|{\bf x}-{\bf x}_s\|_q=\sigma_{s,q}({\bf x}):=\inf_{{\bf y}\in \Sigma_s} \|{\bf x}-{\bf y}\|_q$.
For $q=1$, we use $\sigma_s({\bf x})$  instead of $\sigma_{s,1}({\bf x})$  for brevity.

 In this paper, we introduce a  new property
 of a measurement matrix
 ${\bf A}$:
 there exist positive constants $D$ and $\beta$ such that
   \begin{equation}
 \label{weaksvcondition}
\|{\bf x}_s\|_r^q\le D \|{\bf A} {\bf x}\|_p^q+\beta  s^{q/r-1} (\sigma_{s,q}({\bf x}))^q \quad {\rm for \ all} \ {\bf x}\in {\mathbb R}^n,
 \end{equation}
where $0<p, q, r\le \infty$, $s$ is a positive integer,
 and ${\bf x}_s$ is the best $s$-sparse approximation of the vector ${\bf x}$ in $\ell^q$.
 The property of a measurement matrix mentioned in the abstract
 is a special case of the  above property where $p=r=2$ and $q=1$.
 We call the  property \eqref{weaksvcondition} the {\em sparse approximation property of order $s$},
 as it is closely related to the best $s$-sparse approximation.
We call the minimal constant $\beta$
such that \eqref{weaksvcondition} holds the {\em sparse approximation constant}, and denote it by
 $\beta_{s}({\bf A})$.

 In this paper, we show that
 for the stable recovery of a compressible signal ${\bf x}$
from its noisy measurements ${\bf z}={\bf A}{\bf x}+{\bf n}$ via solving the
$\ell^q$-minimization problem \eqref{minimizationproblem}, the sparse approximation property \eqref{weaksvcondition}
with sparse approximation constant $\beta_{s}({\bf A})<1$
is {\bf sufficient} while the  sparse approximation property \eqref{weaksvcondition}  with finite sparse approximation constant $\beta_s({\bf A})$
 is {\bf necessary}.
 We refer the reader to
 \cite{candesrombergtao06, candes06, cdd09}, \cite{zhang09} -- \cite{sun10}  
  and the references therein
 for other various conditions on a measurement matrix that guarantee the stable recovery of any compressible signal from its
 noisy measurements via solving
 the $\ell^q$-minimization problem \eqref{minimizationproblem}.



 \begin{Tm} \label{maintheorem.tm} Let $0<q\le 1, q\le r\le \infty, 1\le p\le \infty$,
 $ \epsilon\ge 0$,
positive integers $m,n, s$ satisfy $2s\le m\le n$,
  ${\bf A}$ be a matrix of size $m\times n$ having the sparse approximation property
  \eqref{weaksvcondition} with $D\in (0, \infty)$ and $\beta\in (0,1)$,
    ${\bf z}={\bf A}{\bf x}+{\bf n}$ with $\|{\bf n}\|_p\le \epsilon$ and ${\bf x}\in {\mathbb R}^n$,
and let ${\bf x}^*$ be the solution of the $\ell^q$-minimization problem \eqref{minimizationproblem}.
Then
\begin{equation}\label{maintheorem.tm.eq3}
\|{\bf x}^*-{\bf x}\|_{r}^q  \le
\frac{ (3+\beta) D}{1-\beta}   (2\epsilon)^q + \frac{2(1+\beta)^2}{1-\beta} s^{q/r-1} (\sigma_{s,q}({\bf x}))^q
\end{equation}
and
\begin{equation}\label{maintheorem.tm.eq4}
\|{\bf x}^*-{\bf x}\|_{q}^q  \le
\frac{ (3+\beta)D}{1-\beta}  s^{1-q/p} (2\epsilon)^q  + \frac{2(1+\beta)^2}{1-\beta}  (\sigma_{s,q}({\bf x}))^q
\end{equation}
if $q<r$, and
\begin{equation}\label{maintheorem.tm.eq5}
\|{\bf x}^*-{\bf x}\|_{q}^q  \le
\frac{ 2D}{1-\beta}   (2\epsilon)^q + \frac{2(1+\beta)}{1-\beta}  (\sigma_{s,q}({\bf x}))^q
\end{equation}
if $q=r$.
\end{Tm}


\begin{Tm} \label{maintheorem2.tm} Let $ 0<q, p\le \infty $, positive integers
$m,n, s$ satisfy $2s\le m\le n$,
  and let ${\bf A}$ be a matrix of size $m\times n$.
  If for any $\epsilon\ge 0$ and ${\bf x}\in {\mathbb R}^n$, the error
    between the given vector ${\bf x}$ and
   the solution ${\bf x}^*$
of the $\ell^q$-minimization problem \eqref{minimizationproblem} 
satisfies
\begin{equation}\label{maintheorem.tm2.eq1}
\|{\bf x}^*-{\bf x}\|_p^q\le
B_1 \epsilon^q + B_2   s^{q/p-1} (\sigma_{s,q}({\bf x}))^q,
\end{equation}
where
 $B_1$ and $B_2$ are positive constants independent of $\epsilon$ and
${\bf x}$, then
 \begin{equation}
 \label{oldsvcondition}
\|{\bf x}\|_p^q\le B_1 \|{\bf A} {\bf x}\|_p^q+B_2  s^{q/p-1} (\sigma_{s,q}({\bf x}))^q \quad {\rm for \ all} \ {\bf x}\in {\mathbb R}^n,
 \end{equation}
and hence ${\bf A}$ has  the sparse approximation property \eqref{weaksvcondition} with $r=p$, $D=B_1$ and $\beta=B_2$.
%
\end{Tm}


The $m\times n$ adjacency matrix ${\bf \Phi}$ of an unbalanced $(2s, \alpha)$-expander with left degree $d$  and $\alpha\in (0, 1/4)$
satisfies
\begin{equation}\label{expandersap.eq}
\|{\bf x}_s\|_1\le \frac{1}{d(1-4\alpha)} \|{\bf \Phi}{\bf x}\|_1+\frac{2\alpha}{1-4\alpha} \sigma_s({\bf x})\quad {\rm for \ all}\ {\bf  x}\in {\mathbb R}^n,\end{equation}
(and hence it has the sparse approximation property \eqref{weaksvcondition} with $p=q=r=1$). The above property for the adjacency matrix ${\bf \Phi}$ is established in
\cite[Lemma 16]{berinde09} implicitly. Then by \eqref{expandersap.eq} and Theorem \ref{maintheorem.tm}, we have
the following result similar to \cite[Theorem 17]{berinde09}.

\begin{Cr}
\label{expander.cor} Let $\epsilon\ge 0$,  positive integers $m,n, s$ satisfy $2s\le m\le n$,
$\alpha\in (0,1/6)$,
${\bf \Phi}$ be the $m\times n$ adjacency matrix of an unbalanced $(2s, \alpha)$-expander with left degree $d$,
    ${\bf z}={\bf \Phi}{\bf x}+{\bf n}$ with $\|{\bf n}\|_1\le \epsilon$ for some  ${\bf x}\in {\mathbb R}^n$,
and let ${\bf x}^*$ be the solution of the minimization problem \eqref{minimizationproblem} with $p=q=1$.
Then
\begin{equation}\label{expander.cor.eq1}
\|{\bf x}^*-{\bf x}\|_1  \le
\frac{ 4}{ d(1-6\alpha)}   \epsilon+ \frac{2-4\alpha}{1-6\alpha}  \sigma_{s}({\bf x}).\end{equation}
\end{Cr}

The paper is organized as follows.
 One of two  basic properties of a measurement matrix ${\bf A}$ in compressive sampling
 (\cite{donoho03} -- \cite{calderbank10})
 is the
 {\em null space property
  of order $s$} in $\ell^q, 0<q\le 1$; i.e.,
   there exists a positive constant $\gamma$ such that
  \begin{equation}\label{nspcondition.reformulation}
  \|{\bf x}_S\|_q^q\le \gamma \|{\bf x}_{S^c}\|_q^q
  \end{equation}
 hold for all vectors ${\bf x}$ in the null space $N({\bf A})$ of the matrix ${\bf A}$  
 and all sets $S$ with  cardinality $\# S$  less than or equal to $s$.
In Section \ref{nsp.section}, we show  in Theorem \ref{nsptosv.tm0} that
any measurement matrix satisfying  \eqref{weaksvcondition}
 will have the null space property \eqref{nspcondition.reformulation}.
 So the sparse approximation property \eqref{weaksvcondition} of a measurement matrix  can be considered as  a {\bf stronger} version of
 the null space property \eqref{nspcondition.reformulation}.
 The other  basic property of a measurement matrix ${\bf A}$ in compressive sampling  (\cite{candestao05, candesrombergtao06, cdd09},
\cite{donoho03} -- \cite{calderbank10}) is
 the {\em restricted isometry property of order $s$}; i.e.,  there exists a positive constant
$\delta\in (0,1)$ such that
  \begin{equation}\label{ripcondition}
 (1- \delta) \|{\bf x}\|_2^2\le \|{\bf A}{\bf x}\|_2^2\le (1+\delta)\|{\bf x}\|_2^2\quad \ {\rm for \ all} \  {\bf x}\in \Sigma_s.
  \end{equation}
  In Section \ref{rip.section},
we prove that if a measurement matrix   has the restricted isometry property \eqref{ripcondition} of order $2s$ then it has the sparse approximation
 property \eqref{weaksvcondition} with
$p=r=2$, and furthermore the constant $\beta$ in \eqref{weaksvcondition} is small when the restricted isometry constant
 is small, see Theorems \ref{riptosv.tm0} and \ref{riptosv.tm0+} for details. Thus
 the sparse approximation  property \eqref{weaksvcondition} of a measurement matrix  can also thought of as a {\bf weaker} version of the restricted
 isometry property \eqref{ripcondition}, see also 
 Remarks \ref{ripsparse.re} and \ref{ripsparse.re2}.
The proofs of all theorems are included in the appendix.




\section{Null space property and  sparse approximation property}\label{nsp.section}

Let ${\mathcal R}({\bf A})$ be the set of matrices ${\bf R}$  satisfying ${\bf A}={\bf A} {\bf R}{\bf A}$, and denote by $\|{\bf R}\|_{p\to q}$
the operator norm of a matrix ${\bf R}$ from $\ell^p$ to $\ell^q$, i.e.,
$\|{\bf Rx}\|_{q}\le \|{\bf R}\|_{p\to q} \|{\bf x}\|_p$ for all vectors ${\bf x}$.
In this section, we show that any measurement matrix satisfying \eqref{weaksvcondition} will have the null space property \eqref{nspcondition.reformulation}
with its null space constant  less than or equal to the constant $\beta$ in \eqref{weaksvcondition}.
Here {\em null space constant} $\gamma_s({\bf A})$ of a measurement matrix ${\bf A}$ is the minimal constant $\gamma$
such  that  \eqref{nspcondition.reformulation} holds. %

%


 \begin{Tm}\label{nsptosv.tm0} Let $0<q\le r\le \infty$, $0<p<\infty$, 
integers $ m, n, s$ satisfy  $2\le 2s\le m\le n$, and ${\bf A}$ be a matrix of size $m\times n$.
Then the following statements hold.
\begin{itemize}
\item[{(i)}]
 If the matrix ${\bf A}$ has the sparse approximation property \eqref{weaksvcondition},
then
 it has the null space property of order $s$ in $\ell^q$ with
 its null space constant $\gamma_s({\bf A})\le \beta_s({\bf A})$.

\item[{(ii)}] If the matrix ${\bf A}$ has
the null space property of order $s$ in $\ell^q$ 
with the null space constant $\gamma_s({\bf A})$, then
it has the sparse approximation property \eqref{weaksvcondition} with $p=q=r$,
$D=\max(1, \gamma_s({\bf A})) \inf_{{\bf R}\in {\mathcal R}({\bf A})} \|{\bf R}\|_{q\to q}^q$ and $\beta=\gamma_{s}({\bf A})$; i.e.,
\begin{eqnarray*}\label{nsptosv.tm0.eq1}
\|{\bf x}_s\|_q^q \!\!& \le &\!\! \big( \max(1, \gamma_s({\bf A})) \inf_{{\bf R}\in {\mathcal R}({\bf A})} \|{\bf R}\|_{q\to q}^q\big)
 \|{\bf A} {\bf x}\|_q^q\nonumber\\
 & &
 +\gamma_s({\bf A}) (\sigma_{s,q}({\bf x}))^q \quad {\rm for \ all} \  {\bf x}\in {\mathbb R}^n.
\end{eqnarray*}
 \end{itemize}
 \end{Tm}

 Applying Theorems \ref{maintheorem.tm} and \ref{nsptosv.tm0} with $p=q=r=1$, we have the following  result
on recovering compressible signals from noisy measurements
 when the measurement matrix has the null space property of order $s$ in $\ell^1$, which is obtained in \cite{Daubechies10} for the noiseless  case. 

  \begin{Cr} \label{nsp.cr} Let
 $ \epsilon\ge 0$,
$m,n, s$ be positive integers with $2s\le m\le n$,
  ${\bf A}$ be a matrix of size $m\times n$ satisfying
  the null space property
  \eqref{nspcondition.reformulation} with $q=1$,
  ${\bf z}={\bf A}{\bf x}+{\bf n}$ with $\|{\bf n}\|_1\le \epsilon$ and ${\bf x}\in {\mathbb R}^n$,
and let ${\bf x}^*$ be the solution of the minimization problem
\eqref{minimizationproblem}  with $p=q=1$.
 If the  null space constant $\gamma_s({\bf A})\in (0,1)$, then
$$ \|{\bf x}^*-{\bf x}\|_1  \le
\frac{ 4\inf_{{\bf R}\in {\mathcal R}({\bf A})} \|{\bf R}\|_{1\to 1}}{1-\gamma_s({\bf A})}   \epsilon
+ \frac{2+2\gamma_s({\bf A})}{1-\gamma_s({\bf A})}  \sigma_{s}({\bf x}).
$$
\end{Cr}




\begin{re}{\rm
The null space property of a measurement matrix is invariant under preconditioning, i.e., if a measurement matrix ${\bf A}$ has the null space property
\eqref{nspcondition.reformulation} then the preconditioned matrix ${\bf P}{\bf A}$ has the null space property
\eqref{nspcondition.reformulation} with the same null space constants, where a preconditioner
 is a nonsingular matrix ${\bf P}$.
 The sparse approximation property \eqref{weaksvcondition} is weakly preconditioning-invariant
 in the sense that
if a measurement matrix ${\bf A}$ satisfies \eqref{weaksvcondition} then
the preconditioned matrix ${\bf P}{\bf A}$ also satisfies \eqref{weaksvcondition} with  $D$ replaced by $\|{\bf P}^{-1}\|_{p\to p}D$.
 This suggests appropriate preconditioning the measurement matrix (and hence the noisy measurements) before signal recovery from its noisy measurements via solving an $\ell^q$-minimization problem.
 }\end{re}


\begin{re} {\rm Let the matrix ${\bf A}$ of  size $m\times n$ have full rank  $m$ (which is the case in most of compressive sampling problems)
and ${\bf A}= {\bf U} \Sigma {\bf V}^t$ be its singular value decomposition. Here  and hereafter ${\bf x}^t$ stands for the transpose of  a vector or a matrix ${\bf x}$.
Then $\Sigma=(\Sigma^\prime\  {\bf 0})$ for some nonsingular diagonal matrix $\Sigma^\prime$.
Now we define the conventional preconditioned measurement matrix $\tilde {\bf A}$ by
$\tilde {\bf A}={\bf P}{\bf A},$
where
${\bf P}=(\Sigma^\prime)^{-1} {\bf U}^t$. In this case,  ${\bf R}\in {\mathcal R}({\bf A})$ if and only if
${\bf R}={\bf V} \Big(\begin{array}{c} {\bf I}\\  {\bf B} \end{array}\Big)$,
where  ${\bf I}$ is the unit matrix of size $m\times m$ and
${\bf B}$ is an arbitrary matrix of size $(m-n)\times n$.
 Let  ${\bf v}_i,1\le i\le n$,
be the column vectors of the matrix ${\bf V}$. Then the null space $N({\bf A})$  of the matrix ${\bf A}$
is spanned by ${\bf v}_i, m+1\le i\le n$, and the vectors ${\bf v}_i, 1\le i\le m$, form an orthonormal basis for $N({\bf A})^\perp$, the orthogonal complement of the null space $N({\bf A})$ of the matrix ${\bf A}$.
As the set $\{{\bf R} {\bf x}:\ \|{\bf x}\|_1\le 1\}$ is a polygon, the maximal $\ell^1$-norm  of ${\bf R}{\bf x}, \|{\bf x}\|_1\le 1$,
is then attained at some
 vertices.
Thus
  \begin{eqnarray}\label{nsptosv.rem.eq3}
  & &  \inf_{{\bf R}\in {\mathcal R}(\tilde {\bf A})}\|{\bf R}\|_{1\to 1}=
  \inf_{{\bf R}\in {\mathcal R}(\tilde {\bf A})}\max_{1\le i\le m} \|{\bf R}{\bf e}_i\|_1\nonumber\\
 \!\! & = &\!\!
  \inf_{{\bf B}} \max_{1\le i\le m} \|{\bf v}_i- ({\bf v}_{m+1} \ \cdots \ {\bf v}_n) {\bf B}{\bf e}_i\|_1\nonumber\\
\!\!  & = &\!\!
\inf_{{\bf u}\in N({\bf A})}  \max_{1\le i\le m}\|{\bf v}_i- {\bf u}\|_1,
  \end{eqnarray}
  where ${\bf e}_i, 1\le i\le m$, form the standard orthonormal basis of ${\mathbb R}^m$.
 In other words,  the quantity
  $\inf_{{\bf R}\in {\mathcal R}(\tilde {\bf A})}\|{\bf R}\|_{1\to 1}$
  is the same as the  distance of ${\bf v}_i, 1\le i\le m$, from the null space $N({\bf A})$ in $\ell^1$.
  From \eqref{nsptosv.rem.eq3} it follows that
$\inf_{{\bf R}\in {\mathcal R}(\tilde {\bf A})}\|{\bf R}\|_{1\to 1}
\le \max_{1\le i\le m} \|{\bf v}_i\|_1\le n^{1/2}$.
It would be an interesting topic on preconditioning a  measurement matrix ${\bf A}$
with the null space property \eqref{nspcondition.reformulation} such that the quantity
$\inf_{{\bf R}\in {\mathcal R}(\tilde  {\bf A})}\|{\bf R}\|_{q\to q}, 0<q\le 1$, for the preconditioned matrix $\tilde {\bf A}$
is not a large number.
 } \end{re}

\section{Restricted isometry property and sparse approximation property}\label{rip.section}

In this section, we prove that
if a measurement matrix   has the restricted isometry property \eqref{ripcondition} of order $2s$, then it has the
 sparse approximation property \eqref{weaksvcondition} with
$p=r=2$, and 
the sparse approximation constant 
 is small when the restricted isometry constant
 is small. Here the {\em restricted isometry constant} $\delta_s({\bf A})$ of a measurement matrix ${\bf A}$
  is the smallest positive constant $\delta$  that satisfies \eqref{ripcondition}.

  \begin{Tm}
 \label{riptosv.tm0}
Let $0<q\le 1$, positive integers $m,n,s$  satisfy $2s\le m\le n$, and the matrix ${\bf A}$ of size $m\times n$
 have the restricted
isometry property \eqref{ripcondition} of order $2s$ with  restricted isometry constant $\delta_{2s}({\bf A})\in (0,1)$.
Then for all $ {\bf x}\in \RR^n$,
\begin{eqnarray}\label{riptosv.tm0.eq-2}
\|{\bf x}\|_2^2\!\! & \le& \!\!\frac{\sqrt{1+\delta_{2s}({\bf A})}+\sqrt{2\delta_{2s}({\bf A})}} {(1-\delta_{2s}({\bf A}))\sqrt{1+\delta_{2s}({\bf A})}} \|{\bf A}{\bf x}\|_2^2\nonumber\\
\!\!& & \!\!
+
\Big(\frac{\sqrt{1+\delta_{2s}({\bf A})}+\sqrt{2\delta_{2s}({\bf A})}} {1-\delta_{2s}({\bf A})}\Big)^2 \nonumber\\
& & \quad \times \delta_{2s}({\bf A})
s^{1-2/q}\big( \sigma_{s,q}({\bf x})\big)^2
\end{eqnarray}
 and
\begin{eqnarray}\label{riptosv.tm0.eq-1}
\|{\bf A} {\bf x}\|_2^2 \!\! & \le & \!\!
\big(1+\delta_{2s}({\bf A})+\sqrt{2 \delta_{2s}({\bf A})}\big) \|{\bf x}\|_2^2\nonumber\\
\!\! & &\!\! + \big(1+\sqrt{2\delta_{2s}({\bf A})}\big)\delta_{2s}({\bf A})\nonumber\\
 & & \!\! \times
 s^{1-2/q}  (\sigma_{s,q}({\bf x}))^2. 
\end{eqnarray}
\end{Tm}

 \begin{Tm}
 \label{riptosv.tm0+}
Let $0<q\le r\le \infty, 0< p\le \infty$, positive integers $m,n,s$  satisfy $2s\le m\le n$, and
${\bf A}$ be a matrix of size $m\times n$ that has the sparse approximation property \eqref{weaksvcondition}.
Then 
%
%
\begin{equation}\label{riptosv.tm0+.eq1}
\frac{1}{D}\|{\bf x}\|_r^q\le \|{\bf A} {\bf x}\|_p^q \quad {\rm for \ all} \ {\bf x}\in \Sigma_{s},
\end{equation}
and
\begin{equation}\label{riptosv.tm0+.eq2}
\frac{1-\beta}{2D} \|{\bf x}\|_r^q\le \|{\bf A} {\bf x}\|_p^q \quad {\rm for \ all} \ {\bf x}\in \Sigma_{2s}.
\end{equation}

%
%
\end{Tm}

\begin{re}\label{ripsparse.re} {\rm
From  Theorem \ref{riptosv.tm0}, we see that
a measurement matrix with small restricted isometry constant
 will have the sparse approximation property \eqref{weaksvcondition}
with $p=r=2, D$ close to one and  $\beta$ close to zero.
Conversely for $p=r=2$ we obtain from
Theorem \ref{riptosv.tm0+} that
if a measurement matrix ${\bf A}$ has the sparse approximation property
\eqref{weaksvcondition}
with $ D$ close to one and  $\beta$ close to zero, then
  the   first inequality
in the restricted isometry property \eqref{ripcondition}
 holds for some  constant $\delta$ close to $1/2$ {\bf only}. For $p=q=r=1$, the  $m\times n$ adjacency matrices  ${\bf \Phi}$ of unbalanced
 $(2s, \alpha)$-expander with fixed left degree $d$ has the  sparse approximation property \eqref{weaksvcondition}
  with small sparse approximation constant (see \eqref{expandersap.eq})  and the restricted isometry property with respect to $\ell^1$-norm:
  $$(1-C \alpha) \|{\bf x}\|_1\le \|{\bf \Phi} {\bf x}\|_1/d\le (1+ C \alpha) \|{\bf x}\|_1 \quad {\rm for \ all} \ {\bf x}\in \Sigma_{2s}$$
  where $C$  is a positive constant (see \cite[Theorem 1]{berinde09}),
  but it
  does not have the  restricted isometry property \eqref{ripcondition}  when $m/s^2$ is sufficiently small \cite{chander08}.
}\end{re}

\begin{re}\label{ripsparse.re2}{\rm
If a measurement matrix ${\bf A}$ has the restricted isometry property \eqref{ripcondition} with small restricted isometry constant (see \cite{candestao05, donoho06, calderbank10, rauhut08, rudelson08} for examples of such
measurement matrices),  then the preconditioned measurement matrix
${\bf P}{\bf A}$ has the  sparse approximation property \eqref{weaksvcondition}
with $p=r=2$, $D$ close to $\|{\bf P}^{-1}\|_{2\to 2}$ and $\beta$ close to zero
but it  does not have the restricted isometry property \eqref{ripcondition} in general.
This observation may suggest that preconditioning procedure could generate new measurement matrices
for the stable recovery of compressible signals from their noisy measurements.
}\end{re}

\section{Conclusions and final remarks}
In this paper, we introduce the sparse approximation property \eqref{weaksvcondition} of a measurement matrix and show that it is a sufficient and almost necessary condition  that any compressible signal can be stably recovered from its noisy measurements via solving
the $\ell^q$-minimization problem \eqref{minimizationproblem}.

The sparse approximation property \eqref{weaksvcondition} of a measurement matrix with $q\le r$ is a stronger version of
the null space property \eqref{nspcondition.reformulation} with the preconditioning-invariance almost preserved. The sparse approximation property \eqref{weaksvcondition}  with $p=r=2$ and $0<q\le 1$ is
a weaker version of the restricted isometry property \eqref{ripcondition}.
The adjacency matrices of  some unbalanced expanders have the sparse approximation property \eqref{weaksvcondition} with $p=q=r=1$  and small
sparse approximation constant, but they do not have the restricted isometry property \eqref{ripcondition}.
%
A challenging problem is the construction of measurement matrices, other than random matrices \cite{candestao05, donoho06, calderbank10, rauhut08, rudelson08} and adjacency matrices of a graph \cite{berinde09, chander08,  capalbo02, jafarpour09, ba10},
 that have sparse approximation property \eqref{weaksvcondition} with small sparse approximation constant.

%
%

\begin{appendix}

\subsection{Proof of Theorem \ref{maintheorem.tm}}
%

Set ${\bf h}={\bf x}^*-{\bf x}$.
Let $S_0$ be so chosen that
$\|{\bf x}_{S_0^c}\|_q= \|{\bf x}-{\bf x}_{s}\|_q$,
 $S_1$ be the set of indices of the $s$ largest components, in absolute value, of ${\bf h}$ in $S_0^c$, $S_2$ be the set
of indices of the next $s$  largest components, in absolute value, of ${\bf h}$
 in $(S_0\cup S_1)^c$, and so on.
 Then
\begin{equation}\label{maintheorem.tm.pf.eq-1}
\|{\bf A}{\bf h}\|_p 
\le 2\epsilon \quad {\rm and} \quad \|{\bf h}_{S_0^c}\|_q^q \le \|{\bf h}_{S_0}\|_q^q+ 2\| {\bf x}-{\bf x}_{s}\|_q^q
\end{equation}
 by \eqref{minimizationproblem},
 and
  \begin{equation}\label{maintheorem.tm.pf.eq-2}
 \|{\bf h}_{S_j}\|_{\tilde r}\le s^{1/{\tilde r}-1/q} \|{\bf h}_{S_{j-1}}\|_q,  \quad  j\ge 2\end{equation}
 by the construction of the sets $S_j, j\ge 1$, where $q\le \tilde r\le r$. Combining
\eqref{weaksvcondition}
 and \eqref{maintheorem.tm.pf.eq-1}  gives
\begin{equation}\label{maintheorem.tm.pf.eq1}
\|{\bf h}_T\|_r^q\le D (2\epsilon)^q + \beta s^{q/r-1} \|{\bf h}_{T^c}\|_q^q
\end{equation}
for any subset $T$ of $\{1, \ldots, n\}$ with $\# T\le s$.
Applying  \eqref{maintheorem.tm.pf.eq1} with $T$ replaced by $S_0$ and then using  the estimate
\eqref{maintheorem.tm.pf.eq-1} for  $\|{\bf h}_{S_0^c}\|_q^q$,
\begin{equation}\label{maintheorem.tm.pf.eq4}
\|{\bf h}_{S_0}\|_r^q 
 \le
D (2\epsilon)^q+ 2 \beta s^{q/r-1}\|{\bf x}-{\bf x}_{s}\|_q^q
+ \beta s^{q/r-1} \|{\bf h}_{S_0}\|_q^q.
\end{equation}
By H\"older inequality and the property that $\# S_0\le s$,
\begin{equation}\label{maintheorem.tm.pf.eq5}
\|{\bf h}_{S_0}\|_q\le s^{1/q-1/r} \|{\bf h}_{S_0}\|_r.
\end{equation}
Substituting the above inequality into the right-hand side of the inequality \eqref{maintheorem.tm.pf.eq4}
leads to the first crucial inequality:
\begin{equation}\label{maintheorem.tm.pf.eq6}
\|{\bf h}_{S_0}\|_r^q\le \frac{D}{1-\beta} (2\epsilon)^q + \frac{2\beta}{1-\beta} s^{q/r-1} \|{\bf x}-{\bf x}_{s}\|_q^q.
\end{equation}
Combining \eqref{maintheorem.tm.pf.eq-1}, \eqref{maintheorem.tm.pf.eq5} and \eqref{maintheorem.tm.pf.eq6}
yields the second crucial inequality:
\begin{equation}\label{maintheorem.tm.pf.eq7}
\|{\bf h}_{S_0^c}\|_q^q 
\le  \frac{D}{1-\beta} s^{1-q/r}  (2\epsilon)^q+ \frac{2}{1-\beta} \| {\bf x}-{\bf x}_{s}\|_q^q.
\end{equation}
For $r=q$,  the conclusion
\eqref{maintheorem.tm.eq5} follows from \eqref{maintheorem.tm.pf.eq6} and \eqref{maintheorem.tm.pf.eq7}.

Applying
 \eqref{maintheorem.tm.pf.eq1}   with $T$ replaced by $S_1$  yields
  \begin{equation*}\label{maintheorem.tm.pf.eq8}
\|{\bf h}_{S_1}\|_r^q  \le   D (2\epsilon)^q+ \beta s^{q/r-1} \|{\bf h}_{S_1^c}\|_q^q.
\end{equation*}
This together with \eqref{maintheorem.tm.pf.eq-1},
\eqref{maintheorem.tm.pf.eq5}, \eqref{maintheorem.tm.pf.eq6}  and \eqref{maintheorem.tm.pf.eq7} leads to the third crucial  inequality:
\begin{eqnarray}\label{maintheorem.tm.pf.eq9}
\|{\bf h}_{S_1}\|_r^q & \le &  D (2\epsilon)^q+ \beta s^{q/r-1} \big(\|{\bf h}_{S_0}\|_q^q+
\|{\bf h}_{(S_0\cup S_1)^c}\|_q^q\big) \nonumber\\
& \le & \frac{D(1+\beta) }{1-\beta} (2\epsilon)^q + \frac{2\beta(1+\beta)}{1-\beta} s^{q/r-1} \|{\bf x}-{\bf x}_{s}\|_q^q.\nonumber\\
&& \end{eqnarray}
Therefore for $q\le \tilde r\le r$,
\begin{eqnarray}\label{maintheorem.tm.pf.eq10}
\|{\bf h}\|_{\tilde r}^q 
& \le &
\|{\bf h}_{S_0}\|_{\tilde r}^q+ \|{\bf h}_{S_1}\|_{\tilde r}^q+\sum_{j\ge 2} \|{\bf h}_{S_j}\|_{\tilde r}^q\nonumber\\
& \le & s^{q/{\tilde r}-q/r} \|{\bf h}_{S_0}\|_r^q+ s^{q/{\tilde r}-q/r} \|{\bf h}_{S_1}\|_r^q\nonumber\\
& &
+ s^{q/{\tilde r}-1} \|{\bf h}_{S_0^c}\|_q^q\nonumber\\
& \le & \frac{ D(3+\beta)}{1-\beta}  s^{q/{\tilde r}-q/r} (2\epsilon)^q\nonumber\\
 & &
  + \frac{2(1+\beta)^2}{1-\beta} s^{q/{\tilde r}-1} \|{\bf x}-{\bf x}_{s}\|_q^q,
\end{eqnarray}
where the first inequality holds by the triangle inequality for $\|\cdot\|_{q/{\tilde r}}^{q/{\tilde r}}$ as $q\le \tilde r$, the second inequality
is true by H\"older inequality and \eqref{maintheorem.tm.pf.eq-2}, and the third inequality follows from
 \eqref{maintheorem.tm.pf.eq6},  \eqref{maintheorem.tm.pf.eq7} and \eqref{maintheorem.tm.pf.eq9}.
Then the conclusions \eqref{maintheorem.tm.eq3} and \eqref{maintheorem.tm.eq4}
 follow by letting  $\tilde r=r$ and $\tilde r=q$ in \eqref{maintheorem.tm.pf.eq10} respectively.

\subsection{Proof of Theorem \ref{maintheorem2.tm}}
The conclusion
\eqref{oldsvcondition} follows from the estimate
\eqref{maintheorem.tm2.eq1} and the observation that the zero vector is the solution of the $\ell^q$-minimization problem \eqref{minimizationproblem}
with $\epsilon=\|{\bf A}{\bf x}\|_p$ and ${\bf z}={\bf A}{\bf x}$ for any ${\bf x}\in {\mathbb R}^n$.



\subsection{Proof of Theorem \ref{nsptosv.tm0}}
  (i)\quad Take a vector ${\bf x}\in {\mathbb R}^n$ with ${\bf A}{\bf x}={\bf 0}$ and let ${\bf x}_s$ be its best $s$-sparse approximation in
   $\ell^q$.
 Then it is a best $s$-sparse approximation in $\ell^r$. This together with the sparse approximation property \eqref{weaksvcondition}
 leads to
$
\|{\bf x}_s\|_q^q\le s^{1-q/r} \|{\bf x}_s\|_r^q\le \beta_{s}({\bf A})  (\sigma_{s,q}({\bf x}))^q$.
Thus  the measurement matrix ${\bf A}$ has the null space property of order $s$ with $\gamma_{s}({\bf A})\le \beta_{s}({\bf A})$.

(ii)\quad
 Take a vector ${\bf x}\in {\mathbb R}^n$.
 Then
 it suffices to prove that
 \begin{eqnarray}\label{nsptosv.tm0.pf.eq1}
 \|{\bf x}_T\|_q^q   & \le &    \Big(
 \max(1, \gamma_s({\bf A})) \inf_{{\bf R}\in {\mathcal R}({\bf A})}\|{\bf R}\|_{p\to q}^q\Big) \|  {\bf A} {\bf x}\|_p^q \nonumber\\
 & &
 + \gamma_s({\bf A}) \|{\bf x}_{T^c}\|_q^q,
 \end{eqnarray}
for all subsets $T\subset \{1, \ldots, n\}$ with $\# T\le s$,
 where $0<p\le \infty$.
 Note that
 $ {\bf A}({\bf x}- {\bf R}{\bf A} {\bf x})= ({\bf A}-{\bf A} {\bf R} {\bf A}) {\bf x}={\bf 0}$ for   all ${\bf R}\in {\mathcal R}({\bf A})$.
 This together with the null space property \eqref{nspcondition.reformulation} of the measurement matrix ${\bf A}$ leads to
$\|({\bf x}- {\bf R}{\bf A} {\bf x})_T\|_q^q\le \gamma_s({\bf A}) \|({\bf x}- {\bf R}{\bf A} {\bf x})_{T^c}\|_q^q
$ 
for all subsets $T\subset \{1, \ldots, n\}$ with $\# T\le s$ and ${\bf R}\in {\mathcal R}({\bf A})$. Hence
\begin{equation*} \label{nsptosv.tm0.pf.eq4}
\|{\bf x}_T\|_q^q 
\le
 \max(1, \gamma_s({\bf A})) \|{\bf R}\|_{p\to q}^q \|  {\bf A} {\bf x}\|_p^q+ \gamma_s({\bf A}) \|{\bf x}_{T^c}\|_q^q.\quad
\end{equation*}
Taking minimum over all matrices ${\bf R}\in {\mathcal R}({\bf A})$
in the right-hand side of the above estimate leads to
\eqref{nsptosv.tm0.pf.eq1}, and hence proves the second conclusion.

\subsection{Proof of Theorem \ref{riptosv.tm0}}
 Take a vector ${\bf x}\in {\mathbb R}^n$ and let ${\bf x}_s$ be its $s$-sparse approximation in $\ell^2$. We write
${\bf x}=\sum_{j\ge 0} {\bf x}_{S_j}$,
where
 $S_0$ is the set of indices of the $s$ largest components, in absolute value, of ${\bf x}$,
 $S_1$ is the set of indices of the $s$ largest components, in absolute value, of ${\bf x}$  in $S_0^c$,  and so on.
From the construction of the sets $S_j, j\ge 0$, we obtain that ${\bf x}_{S_0}={\bf x}_s$,
$\|{\bf x}_{S_0^c}\|_q=\sigma_{s,q}({\bf x})$,
$\sum_{j\ge 2} \|{\bf x}_{S_j}\|_2 \le
s^{1/2-1/q}\sigma_{s,q}({\bf x})$,
\begin{eqnarray}\label{riptosv.tm0.pf.eq0}
\|{\bf x}_{S_j}\|_2 \!\! & \le & \!\!
s^{1/2-1/q}\|{\bf x}_{S_{j-1}}\|_q^{1-q/2} \|{\bf x}_{S_j}\|_q^{q/2}\nonumber\\
\!\! & \le &\!\!
s^{1/2-1/q}\|{\bf x}_{S_{j-1}}\|_q
\end{eqnarray}
for all $j\ge 1$, and
\begin{eqnarray}\label{riptosv.tm0.pf.eq0+}
\|{\bf A}{\bf x}\|_2^2 & = & \| {\bf A} ({\bf x}_{S_0}+{\bf x}_{S_1})\|_2^2+
\sum_{j\ge 2} \|{\bf A} {\bf x}_{S_j}\|_2^2\nonumber\\
& & + 2
\sum_{j\ge 2} \langle {\bf A} {\bf x}_{S_0}, {\bf A} {\bf x}_{S_j}\rangle + 2\sum_{j\ge 2}\langle {\bf A} {\bf x}_{S_1}, {\bf A} {\bf x}_{S_j}\rangle 
\nonumber\\
& &
+  2
\sum_{2\le j<j'}\langle  {\bf A} {\bf x}_{S_j}, {\bf A} {\bf x}_{S_{j'}}\rangle.
\end{eqnarray}
%
%
Recalling that
$
|\langle {\bf A} {\bf x}_{S_j},  {\bf A}{\bf x}_{S_{j'}}\rangle|\le \delta_{2s}({\bf A}) \|{\bf x}_{S_j}\|_2 \|{\bf x}_{S_{j'}}\|_2
$ for all $ j'\ne j$ (\cite{candestao05}), and applying the restricted isometry property \eqref{ripcondition},
 we obtain from \eqref{riptosv.tm0.pf.eq0} and \eqref{riptosv.tm0.pf.eq0+} that
 \begin{eqnarray*}\label{riptosv.tm0.pf.eq1+}
 (1-\delta_{2s}({\bf A})) \| {\bf x}\|_2^2
\!\! & \le & \!\!
   \|{\bf A}{\bf x}\|_2^2+\delta_{2s}({\bf A}) s^{1-2/q}  (\sigma_{s,q}({\bf x}))^2\nonumber\\
\!\!   & & \!\! + 2 \sqrt{2} \delta_{2s}({\bf A})s^{1/2-1/q} \|{\bf x}\|_2 \sigma_{s,q}({\bf x})\nonumber\\
\!\! & \le &\!\!
   \|{\bf A}{\bf x}\|_2^2 
   +  \delta_{2s}({\bf A}) \epsilon \|{\bf x}\|_2^2\nonumber\\
\!\!   & & \!\! + \delta_{2s}({\bf A}) (1+2\epsilon^{-1}) s^{1-2/q} (\sigma_{s,q}({\bf x}))^2,
   \end{eqnarray*}
   where $\epsilon>0$.
   Then
   \eqref{riptosv.tm0.eq-2} follows by taking $\epsilon=-2+\sqrt{2 (\delta_{2s}({\bf A}))^{-1}+2}$.

Similarly we get 
\begin{eqnarray*}
\|{\bf A}{\bf x}\|_2^2 
 \!\! & \le & \!\! \big(1+\delta_{2s}({\bf A})+\sqrt{2 \delta_{2s}({\bf A})}\big) \|{\bf x}\|_2^2\nonumber\\
 & & + \big(1+\sqrt{2\delta_{2s}({\bf A})}\big)\delta_{2s}({\bf A}) s^{1-2/q}  (\sigma_{s,q}({\bf x}))^2.
\end{eqnarray*}
This proves  \eqref{riptosv.tm0.eq-1}.

\subsection{Proof of Theorem \ref{riptosv.tm0+}}
Take an $s$-sparse vector ${\bf x}\in {\mathbb R}^n$. Then ${\bf x}={\bf x}_s$
 and $\sigma_{s, q}({\bf x})=0$.
This together with the sparse approximation  property \eqref{weaksvcondition}
 gives
 $
 \|{\bf x}\|_r^q=\|{\bf x}_s\|_r^q
\le D \|{\bf A} {\bf x}\|_p^q+ \beta  s^{q/p-1} \sigma_{s, q}({\bf x})^q
 =D \|{\bf A} {\bf x}\|_p^q$,
and hence proves \eqref{riptosv.tm0+.eq1}.

 Take a $2s$-sparse vector ${\bf x}\in {\mathbb R}^n$ and  write
 ${\bf x}={\bf x}_{S_0}+{\bf x}_{S_1}$ for some subsets $S_0$ and $S_1$ of $\{1, \ldots, n\}$
  with empty intersection and cardinality less than or equal to $s$.
  Applying \eqref{weaksvcondition}
    to the given $2s$-sparse vector ${\bf x}$ and replacing $S$  by  $S_0$ and $S_1$ respectively, we obtain
 \begin{equation}\label{svtorip.tm0+.pf.2eq1}
 \|{\bf x}_{S_0}\|_r^q\le D \|{\bf A} {\bf x}\|_p^q+ \beta  s^{q/p-1} \|{\bf x}_{S_1}\|_q^q\le
 D \|{\bf A} {\bf x}\|_p^q+ \beta   \|{\bf x}_{S_1}\|_r^q
  \end{equation}
and
 \begin{equation}\label{svtorip.tm0+.pf.2eq2} \|{\bf x}_{S_1}\|_r^q\le D \|{\bf A} {\bf x}\|_p^q+ \beta  s^{q/p-1} \|{\bf x}_{S_0}\|_q^q\le
 D \|{\bf A} {\bf x}\|_p^q+ \beta   \|{\bf x}_{S_0}\|_r^q.
  \end{equation}
Summing up  the above estimates \eqref{svtorip.tm0+.pf.2eq1} and \eqref{svtorip.tm0+.pf.2eq2}
yields the following  inequality:  
\begin{eqnarray*}
(1-\beta)\|{\bf x}\|_r^q &  = &
(1-\beta)(\|{\bf x}_{S_0}\|_r^r+\|{\bf x}_{S_1}\|_r^r)^{q/r}\nonumber\\
& \le &
(1-\beta)(\|{\bf x}_{S_0}\|_r^q+\|{\bf x}_{S_1}\|_r^q)\le 2 D \|{\bf A} {\bf x}\|_p^q. 
\end{eqnarray*}
Hence \eqref{riptosv.tm0+.eq2} follows.
%
%

%

\end{appendix}


\noindent{\bf Acknowledgement.} {\rm Part of this work was done when the  author  visited
Ecole Polytechnique Federale de Lausanne. 
The author would like to thank Professors   Michael Unser and Martin Vetterli for the hospitality and  fruitful discussion,
and also Professor Namrata Vaswani and anonymous reviewers for providing useful remarks and additional references, and for pointing out that
the adjacency matrices of  unbalanced expanders have the sparse approximation property.}

\begin{biography}
{Qiyu Sun} received the BSc
and PhD degree in mathematics from Hangzhou University, China in
1985 and 1990 respectively. He is currently with Department of
Mathematics, University of Central Florida. His prior position was
with  Zhejiang University (China),  National University of
Singapore (Singapore),  Vanderbilt University, and University of
Houston.
His research interest includes  sampling theory, Wiener's lemma, multi-band
wavelets,  frame theory, linear and nonlinear inverse problems,  and Fourier analysis. He has
published more than 100 papers on mathematics and signal processing,  and written a book ``{\em An Introduction
to Multiband Wavelets}" (Zhejiang University Press, 2001) with Ning Bi and Daren Huang. He is on the editorial board of  the journals ``{\em Advance in Computational Mathematics}" and ``{\em Numerical Functional Analysis and Optimization}".
\end{biography}\vspace{-10mm}

\end{document}